\documentstyle[12pt,epsf]{ioplppt}

\begin{document}

\jl{4}

\title{
\hfill {\normalsize LBNL-40307}\\
\bigskip
Properties of Exotic Matter
for Heavy Ion Searches}[Exotic Matter]

\author{J Schaffner-Bielich\dag, C Greiner\ddag, 
H St\"ocker\S\ and A P Vischer\P}

\address{\dag\
Nuclear Science Division, Lawrence Berkeley National Laboratory, 
University of California, Berkeley, CA 94720, USA}
\address{\ddag\
Institut f\"ur Theoretische Physik,
Justus-Liebig Universit\"at,
D-35392 Giessen, Germany}
\address{\S\
Institut f\"ur Theoretische Physik,
J.W. Goethe-Universit\"at,
D-60054 Frankfurt, Germany}
\address{\P\
Niels Bohr Institute, 
Blegdamsvej 17, 
DK-2100 Copenhagen, Denmark}

\begin{abstract}
We examine the properties of both forms of strange matter, small lumps of
strange quark matter (strangelets) and of strange hadronic matter (Metastable
Exotic Multihypernuclear Objects: MEMOs) and their relevance for present and
future heavy ion searches.
The strong and weak decays are discussed separately to distinguish between
long-lived and short-lived candidates where the former ones are
detectable in present heavy ion experiments while the latter ones in future
heavy ion experiments, respectively.
We find some 
long-lived strangelet candidates which are highly negatively charged 
with a mass to charge ratio like a
anti deuteron ($M/Z \approx -2$) but masses of A=10 to 16.
We predict also many short-lived candidates, both in quark and in 
hadronic form,
which can be highly charged. Purely hyperonic nuclei like the $\Xi\alpha$
($2\Xi^0 2\Xi^-$) are bound and have a
negative charge while carrying a positive baryon number.
We demonstrate also that multiply charmed exotics 
(charmlets) might be bound and can be produced 
at future heavy ion colliders.
\end{abstract}

\section{Introduction}

Heavy ion collisions offer an unique possibility to study the properties of
hitherto unknown domains of strongly interacting matter.
New forms of matter might be 
possible \cite{bodmer} and formed during the collision.
Strange particles are abundantly produced in central heavy ion collisions 
at relativistic energy. 
This opens up the tantalizing scenario of the formation
of strange matter either by a quark-gluon plasma 
\cite{carsten} or by the coalescence of hyperons \cite{raffa}.
After formation, the system cools down by evaporating baryons and pions via
strong interactions (strong decay). At timescales of $10^{-10}-10^{-5}$~s, 
the system can decay weakly by emitting a baryon or pion and losing one unit
of strangeness (weak hadronic decay). Weak semileptonic decay (emission of
electrons and antineutrinos) will appear then
at a longer time, maybe $10^{-4}$ s after the reaction, as it is a three body
decay \cite{henning}. Most heavy ion experiments searching for strange
matter are sensitive to a lifetime of $\tau \approx 50-100$ ns 
\cite{majka},
i.e.\ they can probably 
see strange matter which is stable against weak hadronic decay (long-lived
candidates) but not the ones which are only stable against strong decay
(short-lived candidates).

In any case, small baryon numbers are expected for the surviving finite
multiply strange objects. 
Hence, shell effects will be important.
Two different classes of strange nuggets are
possible: either a bag consisting of up, down and strange quarks (strangelets)
\cite{farhi} or a 'nucleus' consisting of nucleons and many hyperons or even of
hyperons alone (Metastable Exotic Multihypernuclear Objects, MEMOs)
\cite{scha92}. 
The former ones are calculated by using the MIT bag model with shell mode
filling, the latter ones by using an extended relativistic mean field model.
For an overview of the properties of strange matter for heavy ion physics see
\cite{overview}. 

In the following two sections, 
we discuss the properties of both forms of strange matter 
and the possible long- and short-lived candidates referring to \cite{scha97}.
In the last section, we give an outlook for charmlets at future heavy ion
colliders \cite{axel}.

\section{Long-lived candidates: strangelets}

Up to now, strange quark matter and strangelets have been studied using the MIT
bag model. Whether or not strangelets exist depends crucially on the value of
the bag constant which is not known for such strange and big systems.
For a bag constant of $B^{1/4}=145$ MeV, the original value of the MIT bag
model fit, strangelets are absolutely stable, for bag constants up to $B\approx
180$ MeV strangelets are metastable, i.e.\ they can decay by weak interactions,
for higher bag constants, as suggested by QCD sum rules or fit to charmonium
states, strangelets are unbound.
So anything between absolutely stable and unbound is possible.
Nevertheless, for the following arguments one needs only three basic
assumptions: 
\begin{enumerate}
\item 
Strange quark matter is at least metastable.
\item 
There exists a local minimum for the total energy per baryon 
of strange quark matter at a finite strangeness fraction $f_s=|S|/A$.
\item 
The relativistic shell model can be used for strangelets.
\end{enumerate}
With these assumptions we predict
that there exists a valley of stability at low mass numbers
and that these strangelets are
highly negatively charged contrary to former findings.

\begin{figure}[t]
\epsfysize=0.4\textheight
\centerline{\epsfbox{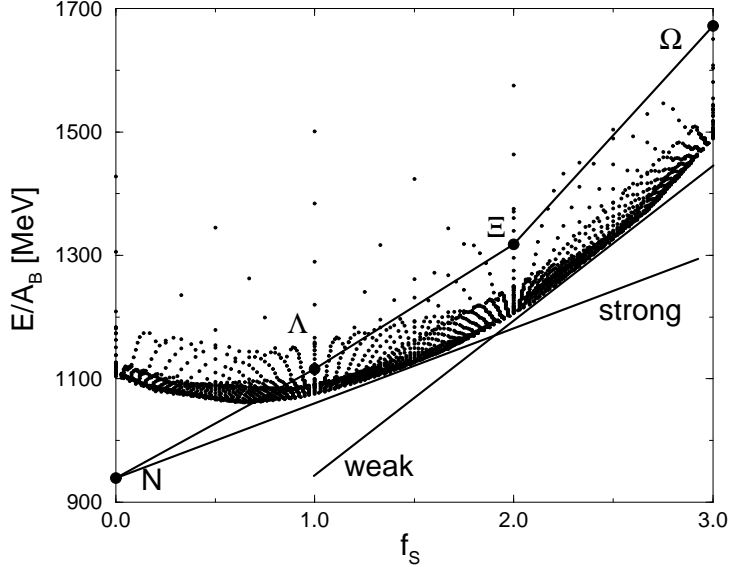}}
\caption{The energy per baryon $E/A_B$ of isospin symmetric strangelets
with $A_B\le40$ for a bag constant of $B^{1/4}=170$ MeV versus
the strangeness fraction $f_s$. The solid line
connects the masses of nucleon, $\Lambda$, $\Xi$ and $\Omega$ and stands
for free baryon matter.}
\label{fig:qsbags}
\end{figure}

The MIT bag model is used here as a guideline only.
Fig.\ \ref{fig:qsbags} shows the energy per baryon number of isospin symmetric 
strangelets as a function of $f_s$ for $A\leq 40$ for a bag parameter of
$B^{1/4}=170$ MeV. Now there are three different processes which will shift a
strangelet emerging from a heavy ion collisions to a very high strangeness
fraction. First, the strangelets sitting above the line drawn between the
nucleon and the hyperon masses will decay to a mixture of nucleons and
hyperons by strong interactions completely as this is energetically
favored. Second, the strangelets located between that line and the tangent
construction starting at the nucleon mass (denoted as strong) can decay
strongly by emitting nucleons and hyperons. They will be shifted to a higher
strangeness fraction until they reach the tangent point at $f_s\approx 1.4$.
Third, weak nucleon decay can occur for the strangelets between the former
tangent and the other tangent (denoted as weak) 
starting at the nucleon mass and $f_s=1$ (as weak
interaction change one unit of strangeness) \cite{chin}. 
For a strangelet with $f_s>1$ 
the weak nucleon decay will enhance the strangeness fraction as 
\begin{equation}
\Delta f_s = \frac{|S|-1}{A-1}- \frac{|S|}{A} = \frac{f_s -1}{A-1}
\quad .
\end{equation}
Hence, strangelets surviving strong and weak
nucleon decay can be sitting at a very high strangeness fraction of
$f_s\approx 2.2$ which is the weak tangent point in Fig.\ \ref{fig:qsbags}.
For isospin symmetric systems, this large strangeness fraction corresponds to a
charge fraction of 
\begin{equation}
 \frac{Z}{A} = \frac{1}{2} \left(1- f_s\right) = -0.55
\end{equation}
which indicates highly charged strange quark matter. This is contrary to the
conventional picture that strangelets have a slightly positive 
charge-to-mass ratio which is the case for strange matter sitting in the
minimum of the curve plotted in Fig.\ \ref{fig:qsbags}.
But as pointed out before,
the combined effect of strong and weak hadronic decay will shift strangelets
emerging from a heavy ion collision to much higher values of $f_s$ and
therefore to highly negatively charged objects! This was indeed also seen in a
dynamical calculation where hadrons were evaporated from a quark-gluon plasma
droplet \cite{carsten2}.

This simplified 
picture is only valid in bulk matter. For finite systems, which we
are interested in, shell effects will be important. Already in Fig.\
\ref{fig:qsbags} one sees that shell effects are at the order of 100 MeV per
baryon number! Hence, we expect that strangelets with a closed shell
can be very deeply bound. These 'magic' strangelets are 
most likely to be stable against
strong and weak hadronic decay modes as their decay products have a much higher
total mass. 
The single particle levels inside a cavity (as for the MIT bag model) or for
ordinary nuclei or hypernuclei show the same order of levels for the lowest
eigenstates. First, there is a 1s$_{1/2}$ shell, then the 1p$_{3/2}$ 
and the 1p$_{1/2}$ shells follow. 
Due to relativistic effects, the spin-orbit splitting is quite sizable
for nucleons. As the quarks are much lighter and relativistic effects are even
more pronounced, the spin-orbit splitting for quarks is at the order of 100
MeV for very light bags, i.e.\ on a similar scale as the splitting between the
s and p shell. 
One can put 6 quarks in the s-shell due to the color degree of freedom, then
12 quarks in the 1p$_{3/2}$ shell and again 6 quarks in the 1p$_{1/2}$ shell.
The smallest and most pronounced 
magic numbers for quarks are then 6, 18, and 24 (the next one would be already
at 42).

Studying isospin asymmetric systems reveals another important effect.
The weak nucleon decay by emitting a proton carries away positive
charge. Nevertheless, the neutron does not carry away negative charge if it is
not accompanied by a $\pi^-$. But this decay is suppressed by the mass of the
pion and the phase space of the three body final state. Therefore, a strangelet
stable against weak nucleon decay is most likely to be negatively charged.

Let us look now for strangelets which have closed shells for all three quark
species with a negative charge 
and a high strangeness fraction as these are the most likely
candidates. The first magic strangelet is the quark alpha
with 6 quarks of each quark species at $A=6$ which has zero charge
\cite{qalpha}. 
The magic strangelets with a high strangeness fraction and a negative
charge are then at $A=10$, $Z=-4$ (with 6 up, 6 down and 18 strange quarks),
$A=12$, $Z=-6$ (with 6 up, 6 down and 24 strange quarks),
$A=14$, $Z=-8$ (with 6 up, 18 down and 18 strange quarks), and
$A=16$, $Z=-10$ (with 6 up, 18 down and 24 strange quarks).
One sees a correlation, that 
adding two units of baryon number decreases the charge by two. 
Note that these strangelets
have a rather high and negative charge fraction of $Z/A
\approx -0.5$ very similar 
to an antideuteron but with a much higher mass and charge! 
These strangelets constitute a valley of stability which is due to 
pronounced shell effects. 

This picture holds, i.e.\ these candidates remain, 
also within an explicit calculation using the MIT bag model
with shell mode filling \cite{scha97}. 
We calculated the masses of strangelets with 
all possible combinations of up, down and strange quarks up to a baryon number
of $A=30$. Then we look for possible strong decays as the emission of 
baryons (p,n,$\Lambda,\Sigma^-,\Sigma^+,\Xi^-,\Xi^0,\Omega^-$) and mesons
(pions and kaons) by calculating the mass difference between the strangelet and
its possible decay products. For the strong interactions, we also allow for
multiple hadron emission, like the strong decay of a strangelet via a neutron and
a pion, and the complete evaporation to hadrons.
For example, the strong proton decay $Q'\to Q+{\rm p}$ is checked by
\begin{equation}
M(A,S,Z) < M(A-1,S,Z-1) + m_p
\end{equation}
where $M(A,S,Z)$ stands for the mass of the strangelet for a given baryon
number, strangeness and charge.
Afterwards we check for weak hadronic decay, the single emission of baryons and
mesons within the same procedure simply by changing one unit of strangeness in
the final products.
The weak proton decay $Q'\to Q+{\rm p}$ is now checked by
\begin{equation}
M(A,S,Z) < M(A-1,S\pm 1,Z-1) + m_p
\end{equation}
where we allow for both strangeness changing processes of $\Delta S=\pm 1$.
This calculation has been done for several bag parameters. We choose a strange
quark mass of $m_s=150$ MeV if not otherwise stated. The value of $B^{1/4}=145$
MeV and $m_s=280$ MeV is taken from the original MIT bag model 
fit to the hadron masses.

\begin{figure}[t]
\epsfysize=0.4\textheight
\centerline{\epsfbox{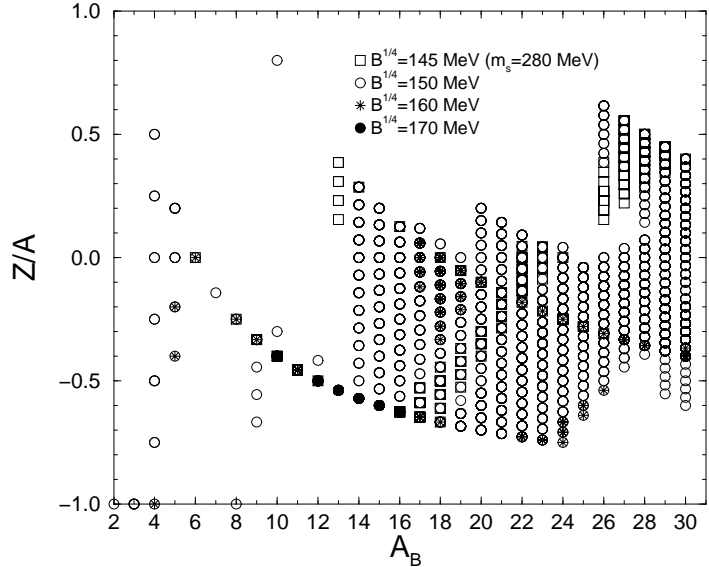}}
\caption{The charge fraction $Z/A$ 
for long-lived strangelets, which are stable against 
strong and weak hadronic decay, for different choices of the bag parameter.
The case for the original MIT bag model 
parameters ($B^{1/4}=145$ MeV, $m_s=280$
MeV) is also plotted.}
\label{fig:stab30}
\end{figure}

The candidates which are stable against strong and weak hadronic decay 
are 
plotted in Fig.\ \ref{fig:stab30} in a scatter plot as a function of their
baryon number and charge fraction.
In all the parametrizations shown, we find the candidates at
$A=10$ with $Z=-4$,
at $A=12$ with $Z=-6$, and at $A=16$ with $Z=-10$.
We do not find any candidates for a bag parameter of $B^{1/4}=180$ MeV 
or higher as strange quark matter starts to get unstable.

As expected and outlined before, 
the main strangelets stable against strong and weak decay are lying
in the valley of stability and are highly negatively charged.
This finding is contrary to the common belief that strangelets have a small
positive 
charge and will have serious impact on present heavy ion searches for strange
matter. In principle, these experiments are able to measure these highly
charged candidates also, but have focussed so far on candidates with a small
charge and/or a high mass \cite{ken,sonja}.

Note, that this calculation does not include colormagnetic and colorelectric
interactions between the quarks.
These interactions have been studied for the candidates at $A\leq 6$ in the
s-shell only \cite{aerts}. 
It was found, that the colormagnetic interaction is mainly
repulsive and results in unbound systems. 
Especially the quark alpha was found to be unbound by 0.9 GeV. 
The only exception is the H dibaryon which is slightly bound when including the
colormagnetic term. These corrections might also change then the overall
picture at $A>6$. But this will only be the case if the corrections are 
larger than the shell effects of about 100 MeV.

\section{Short-lived candidates: MEMOs}

Going back in the timescale of an heavy ion reaction as outlined in the
introduction, one comes to the domain of short-lived strange matter which lives
as short as the hyperons ($\tau\approx 10^{-10}$ s). Multiply strange nuclear
systems can be formed by coalescence of hyperons after a heavy ion
collision \cite{raffa}. 
Indeed, we demonstrated within an extended relativistic mean field
model that MEMOs might exist \cite{scha92} and that
they are even more bound than ordinary nuclei due to the strongly
attractive interaction between the hyperons \cite{scha93}. 
Nevertheless, the hyperon
potentials are not high enough to overcome the mass difference to the nucleons.
Hence, MEMOs can decay weakly on the timescale of the free hyperon weak decay
and are short-lived. Of course, this picture will change if the hyperon-hyperon
interaction is strong enough to create a local minimum in the total energy 
per baryon at large strangeness
fraction which can not be ruled out by our present poor knowledge of multi
hypernuclear properties. 

MEMOs have quite distinct properties, they can be negatively charged while
carrying a positive baryon number due to the negatively charged hyperons, the
$\Sigma^-$ and the $\Xi^-$. 
There exists certain classes of MEMOs: Pauli-blocked systems consisting of 
\{p,n,$\Lambda,\Xi^0,\Xi^-$\} baryons, mixed nucleon and hyperon systems
of e.g.\ \{n,$\Sigma^-,\Xi^-$\} or \{n,$\Lambda,\Xi^-$\} baryons, and
purely hyperonic matter of \{$\Lambda,\Xi^0,\Xi^-$\} baryons.
Very exotic candidates like the alpha particle in the hyperon world, the
$\Xi\alpha$ with two $\Xi^0$ and two $\Xi^-$, have been predicted to be bound. 
Other light candidates are the combinations
$\{2{\rm n},2\Lambda,2\Xi^-\}$,
$\{2{\rm p},2\Lambda,2\Xi^0\}$, $\{2\Lambda,2\Xi^0,2\Xi^-\}$.
Pauli-blocked candidates like 
$^{~~~~6}_{\Xi^0\Xi^0}$He and
$^{~~~~7}_{\Lambda\Lambda\Xi^0}$He.
are discussed in \cite{scha94}.

\begin{figure}[t]
\epsfysize=0.4\textheight
\epsfxsize=\textwidth
\centerline{\epsfbox{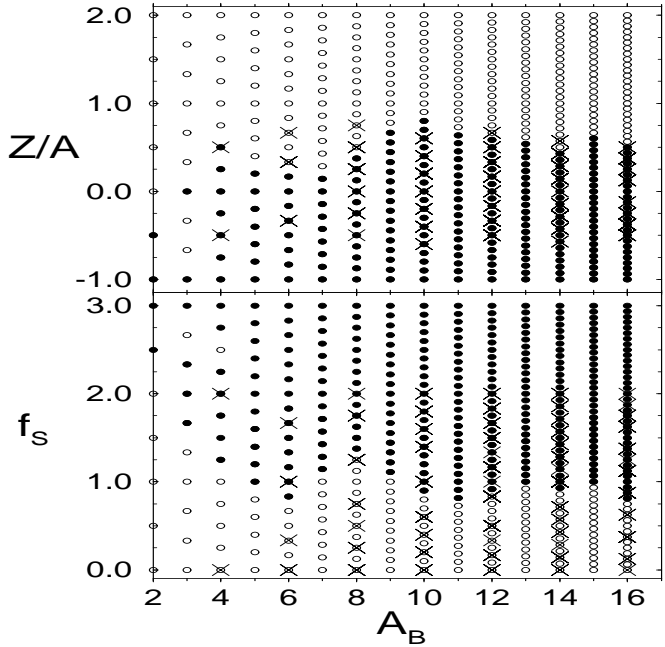}}
\caption{The strangeness per baryon $f_s$ (lower part) and the charge
fraction $Z/A$ (upper part) as a function of the baryon
number $A_B$ for short-lived strangelets (dots) and
unstable strangelets (open circles) for a bag constant of
$B^{1/4}=160$ MeV. The hadronic counterparts, MEMOs, are shown by crosses.}
\label{fig:B160}
\end{figure}

MEMOs compete with strangelets as they are of similar strangeness content.
We calculated light MEMOs up to a closed p-shell and checked for
metastability (strong decay). We analyzed the strangelet candidates without the
weak hadronic decay, i.e.\ allowing for the strong decay only.
The short-lived candidates for MEMOs and 
strangelets for a bag constant of $B^{1/4}=160$ MeV 
are shown in Fig.\
\ref{fig:B160} in a scatter plot as a function of strangeness fraction $f_s$,
charge fraction $Z/A$ and baryon number $A$. 

As can be seen, 
there are many more short-lived candidates than long-lived.
Light MEMOs can have very unusual charge fractions 
between $Z/A=\pm 0.6$ indicating a rich structure of strange hadronic matter.
Strangelet candidates also cover a wide range of charge
fraction but are mainly located at negative charge. This comes from the strong
decay which shifts strangelets to higher strangeness fraction and to negative
charge. 
There are MEMOs and strangelets with the same 
strangeness content and baryon number.
Here, the energetically least favourable object
can decay into the other via strong interactions.
A strangelet created
in a quark gluon plasma can then possibly decay into a MEMO.
Or vice versa, MEMOs can coalesce from the hot and hyperon-rich
zone of a relativistic heavy ion collision first and then they form
a strangelet. 

Presently, there are only experiments designed to look for long-lived
composites with a lifetime of $\tau> 50$ ns 
except for the H dibaryon searches (see e.g.\ \cite{ken,sonja}).
Designing an experiment for short-lived composites is challenging but 
planned for future colliders \cite{coffin} and can reveal
the possibly rich structure of strange matter.

\section{Outlook: charmlets}

With the advent of heavy ion colliders, a new degree of freedom opens: charm.
About ten charm-anticharm pairs are expected in a central collision of gold
nuclei at $\sqrt{s}=200$ GeV at the Relativistic Heavy Ion Collider (RHIC) 
in Brookhaven \cite{ramona}.
In the following we discuss briefly the properties of multiply charmed quark
bags and their production possibility as outlined in \cite{axel}.

It is known from charmonium spectra, that charm quarks have a quite large and
attractive interactions which comes from the Coulomb-like 
colorelectric term of the one-gluon exchange potential.
Indeed, the colorelectric 
term gets stronger with increasing quark mass while the
colormagnetic term decreases with the quark mass. 
Note, that the overall one-gluon exchange in bulk 
is repulsive for massless quarks but gets attractive for heavy quarks.

Another effect is that the colormagnetic term for light quarks (here up, down
and strange quarks) can be enhanced by the presence of charm quarks.
As the light quarks do not need to be in a colorless state anymore, 
a preferred combination of color and spin can be found which has a more
attractive colormagnetic interaction. For example, a system of one up, down and
strange quark (uds)
can now be in a color octet state for a charmlet like \{udsccc\}.
Compared to a color singlet state (the $\Lambda$), 
the colormagnetic term can be
more attractive by 110 MeV in the SU(3) flavor symmetric case \cite{jaffe}
(note that the one-gluon exchange interaction energy is --14 for the
flavor singlet, not --65/3 as given in the table by Jaffe).

\begin{figure}[t]
\epsfysize=0.4\textheight
\centerline{\epsfbox{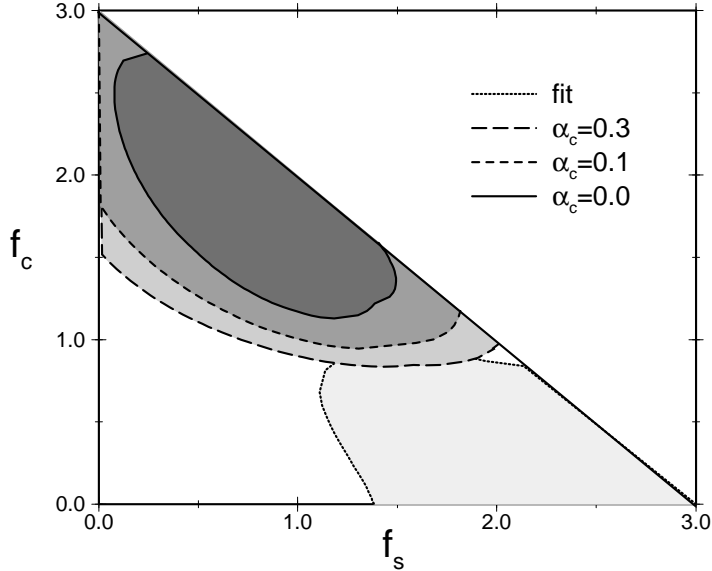}}
\caption{The area of bound strange and charmed matter as a function of
strangeness fraction $f_s=|S|/A$ and charm fraction $f_c=|C|/A$ for various bag
parameters. The case for a fit to the hadron spectra including charmed hadrons
is also shown and is the shaded area at the lower right side.}
\label{fig:bounds}
\end{figure}

Fig.\ \ref{fig:bounds} shows the area of bound charmed strange matter with
respect to hadron emission by strong interactions as a
function of strangeness fraction $f_s$ and charm fraction $f_c=|C|/A$.
The bag parameter has been chosen to be $B^{1/4}=235$ MeV, i.e.\ pure strange
matter is unbound. Still one finds a bound area of charmed strange matter which
increases when increasing the strong coupling constant $\alpha_c$ for the
one-gluon exchange. 

We modified the 
bag model to include heavy quarks by including the colorelectric term in the
same way as the colormagnetic one and are able to fit the 
masses of the hadrons including charm on the level of a few percent. 
The area of bound charm strange matter
for this case is also shown and denoted as fit. Nevertheless,
the bag model gives such a high $\alpha_c$ coupling constant, i.e.\ larger than
$\pi$/8, that the 
one-gluon exchange is a nonperturbative correction
and the pressure for the massless quarks gets negative. Hence,
the bag model
parameters fitted to hadrons can not be applied for bulk matter, also for
charmed matter.

Furthermore, we studied finite systems of multiply charmed exotics.
The binding energy 
is calculated for colormagnetic and colorelectric interactions up to $A=4$
(see \cite{axel} for details).
We find, that charmlets can be bound by 100 to 200 MeV in 
SU(3) flavor symmetry. Their charges ranges from $Z=-4$ to $Z=+4$, again they
can be highly charged. 
The production of charmlets is estimated by using a coalescence model
and the momentum distribution of charm quarks using the HIJING 
model. 
The production rates for double charmed 
(about $10^{-2}$ per event) and maybe even
for triple charmed ($10^{-5}$ per event) are high enough to be seen at future
experiments at RHIC. The main challenge here is again the lifetime of 
charmlets:
as they are so heavy, they decay on the lifetime of the charmed
hadrons, i.e.\ $\tau\approx 10^{-13}$ s. 
A silicon vertex detector is needed to possibly detect these charmed exotics
which will be available in the future 
at the STAR detector at RHIC and at the detector ALICE
at the Large Hadron Collider (LHC) in CERN \cite{coffin}.

\ack
J.S.B. acknowledges support by the Alexander von Humboldt-Stiftung with a
Feodor-Lynen fellowship and thanks all the strangelet searchers for the
numerous discussions during this conference. J.S.B. thanks also
Ziwei Lin for calculating the momentum distribution of charm quarks
using the HIJING model which was used as input for the coalescence calculation.
This work was supported in part by GSI, BMBF and DFG and 
the Director, Office of Energy Research,
Office of High Energy and Nuclear Physics, Nuclear Physics Division of the
U.S. Department of Energy under Contract No.\ DE-AC03-76SF00098.

\end{document}